\documentclass[nofootinbib, reprint, amsmath,amssymb, aps,]{revtex4-2}
\usepackage{graphicx}
\usepackage{dcolumn}
\usepackage{bm}
\usepackage[colorlinks]{hyperref}
\usepackage{soul}

\begin{document}

\title{Does Reconnection Change Magnetic Topology?}


\author{Amir Jafari}
\email{elenceq@jhu.edu}

\begin{abstract}

{\color{black}We employ well-known concepts from statistical physics, quantum field theories and general topology to study magnetic reconnection, topology-change and their connection in incompressible flows in the context of an effective field theory without appealing to magnetic field lines. We consider the dynamical system corresponding to wave-packets moving with Alfv\'en velocity $\dot{\bf x}(t):={\bf V}_A({\bf x},t)$ whose trajectories ${\bf x}(t)$ define path lines, which naturally provides a mathematical way to estimate the rate of magnetic topology-change. A considerable simplification is attained, in fact, by directly employing well-known concepts from hydrodynamic turbulence without appealing to the complicated notion of magnetic field lines moving through plasma, which may prove even more useful in the relativistic regime.} Continuity conditions for magnetic field allow rapid but continuous divergence of path lines, i.e., shown to imply reconnection, but not discontinuous divergence which would change topology. Thus topology can change only due to time-reversal symmetry breaking e.g., by dissipative effects. In laminar and even chaotic flows, the separation of path lines at all times remains proportional to their initial separation, argued to correspond to slow reconnection, and topology changes by dissipation with a rate proportional to resistivity. In turbulence, path lines diverge super-linearly with time \textit{independent} of their initial separation, i.e., fast reconnection, and magnetic topology changes by turbulent dissipation with a rate \textit{independent} of small-scale plasma effects. 
 The crucial role of turbulence in enhancing topology-change and reconnection rates originates from its ability to break time-reversal invariance and make the flow \textit{super-chaotic}. In fact, due to the loss of Lipschitz continuity of magnetic field in turbulence, path lines separate super-linearly even if their initial separation tends to vanish, unlike deterministic chaos. This super-chaotic behavior is an example of spontaneous stochasticity in statistical physics, sometimes called the real butterfly effect in chaos theory to distinguish it from the butterfly effect in which trajectories can diverge exponentially only if initial separation remains finite. If $3D$ reconnection is defined as magnetic topology-change, it can be fast only in turbulence where both reconnection and topology-change are driven by spontaneous stochasticity, independent of any plasma effects. Our results strongly support the Lazarian-Vishniac theory of turbulent reconnection.

\end{abstract}
\pacs{Valid PACS appear here}
\maketitle
\section{Introduction}

In astrophysics, magnetic reconnection has been invoked both as a main mechanism that governs the underlying dynamics at large scales, e.g., in launching outflows in stars and accretion disks, and also as an accompanying process working in the background. It has been suggested, as an example for the latter, that reconnection regulates the structure of a strongly magnetized corona \citep{Uzd2008}. Reconnection is widely thought to be intimately related to the magneto-rotational instability (MRI) \cite{Chan1960, Bal1998} and Parker-Rayleigh-Taylor instability \citep{Par1966}, which might in turn interfere or enhance the reconnection rate \citep{Kad2018}. Reconnection may also affect the saturation rate of the MRI and also the generation of nonthermal particles \citep{Hoshi2013}. It is now widely believed that astrophysical reconection (i) is a ubiquitous process occurring in different systems from the solar surface to highly conducting accretion disks; (ii) is fast, i.e., with a rate orders of magnitude faster than the resistive {\color{black} rate}; (iii) is responsible for a variety of other phenomena such as particle acceleration and plasma heating; and in astrophysical systems; and (iv) proceeds in turbulence {\color{black}as astrophysical flows are mostly turbulent.} For a recent review covering these aspects in details see e.g., \cite{Review2020}.  

Magnetic reconnection seems to be ubiquitous in astrophysics, with a vast literature (see \cite{HesseCassak2020,Pontin2022} for a recent review). Yet, there seems to be no consensus on its definition and its relation to magnetic topology change. In fact, sometimes hand-waving arguments based on magnetic field lines in $2D$ setups seem to be generalized to $3D$ without any mathematical or physical justification. {\color{black}
For example, reconnection is sometimes understood or defined as a magnetic topology change. However, this notion, which seems to originate from 2D setups where the field line connecting plasma element A to another element B at time $t_0$, connects A to C but not B at a later time $t_1$, is inadequate to describe 3D configurations.In 2D reconnection, there is necessarily a discontinuous change in magnetic connectivity implying topology change (leading to different regions with different "topologies" separated by separatrices). In 3D, on the other hand, this process can proceed without any discontinuous change in magnetic connectivity (thus without topology change and no separatrices) because there is "enough" space for field lines to flip \cite{Priestetal95}.
}

Our aim in this paper is to present a formalism to understand reconnection, magnetic topology and topology-change without invoking the concept of magnetic field lines, their motion through plasma or the notion of flux-freezing in laminar \citep{Alfven1942} or turbulent flows \citep{Eyink2011}. These concepts have of course been greatly reformulated and refined in the last decades, in particular for turbulent flows (e.g., see \cite{Eyink2011,Eyink2015,Eyinketal2013,Lazarianetal2015,Lalescuetal2015}). However, a much simpler picture seems to arise using magnetic path-lines, instead of field lines, which is what we will do in this paper. {\color{black}This approach may also prove very useful in relativistic regime, where  appealing to the notion of magnetic field lines becomes even more problematic.} Consider small magnetic disturbances or wave packets moving along the local magnetic field with Alfv\'en velocity ${\bf V}_A$.\footnote{
This is motivated by the concept of quasiparticles in quantum field theories, where e.g., vibrational modes in a crystal are taken as {\color{black}quasiparticles (phonons). Our approach is however much simpler here as we treat these entities as classical Lagrangian particles in fluid approximation.}} {\color{black}These magnetic excitations or} "Alfv\'enic wave-packets" will be taken as fluid particles in a Lagrangian description, i.e.,  we will consider wave-packets as particles moving with Alfv\'en velocity and study their trajectories and the corresponding topology.

\begin{figure}[t]
 \begin{centering}
\includegraphics[scale=.4]{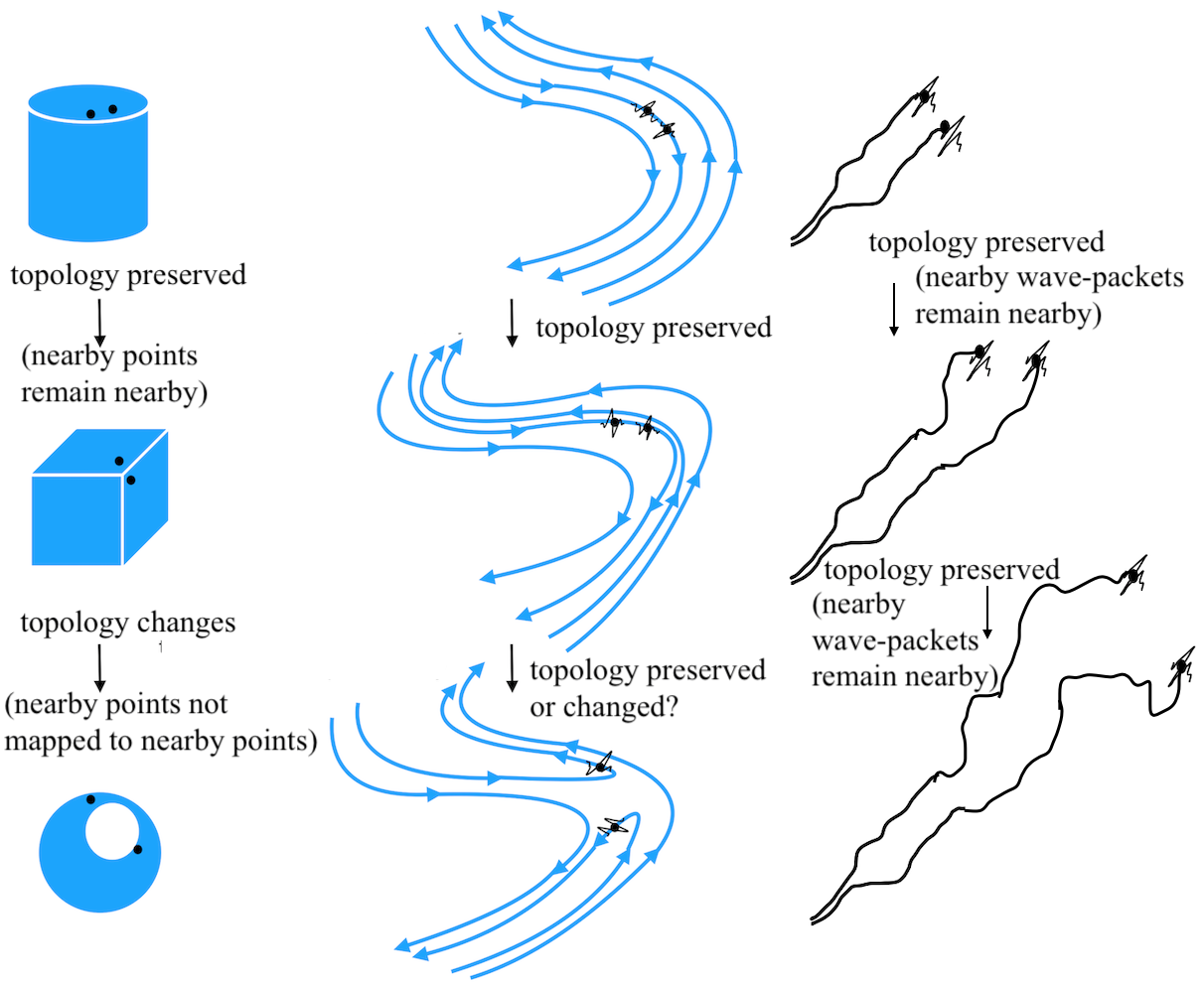}
\caption {\footnotesize {{\color{black}Reconnection vs. magnetic topology change. Left: the topology of an object, e.g., a solid cylinder, is preserved under a deformation as long as it involves only stretching and bending but not cutting and gluing. Mathematically, such a deformation translates into a continuous mapping: nearby points are mapped onto nearby points. Since such a deformation must be reversible, the map must have a continuous inverse. Also, every point must be mapped to one and only one point (no point being destroyed), i.e., the map must be one-to-one and onto. Hence, topology is preserved under such continuous, one-to-one and onto maps with a continuous inverse (i.e, homeomorphisms). A solid cylinder for example is homeomorphic to a solid cube but not to a ball with a hole in it, making which requires cutting.
Middle: Instead of deforming geometrical objects, consider time evolution of magnetic field and instead of points on objects, consider wave-packets, moving along the field ${\bf B(x},t)$ with the local Alfv\'en velocity, $d{\bf x}(t)/dt={\bf V}_A({\bf x}(t), t)$. Magnetic topology is preserved if  initially nearby magnetic wave-packets remain nearby at a slightly later time. This provides an intuitive topology for a continuous field, which would change only due to dissipation (in the presence of which, the mapping is not one-to-one and onto anymore due to time-reversal symmetry breaking). Right: Rate of continuous separation of Alfv\'enic trajectories determines reconnection rate; in non-turbulent flows, the rate depends on initial separation (slow) but in turbulence it is super-linear in time and independent of initial separation (fast). Topology change would correspond to (i) discontinuous divergence of wave-packets (not allowed by continuity of $\bf B$) and/or (ii) resistive dissipation or turbulence which break time-reversal invariance.}  }}\label{topology1}
\end{centering}
\end{figure}

Magnetic field lines provide a powerful notion in many problems, nevertheless, their behavior in real plasmas seem to be complicated enough that other simpler approaches could be appreciated. At any given time $t_0$, magnetic field lines are defined as parametric curves, e.g., $\xi(s;t_0)$ with arc-length parameter $s$, which provide a "pattern" for the field in real space at a given time, i.e, a "snapshot" of the field. {\color{black}In real astrophysical fluids, which are turbulent, field lines become stochastic and the notion of a single field line loses its meaning unless a proper coarse-graining is applied, i.e., field lines of the average, large-scale field are considered \citep{Lazarianetal2015,Review2020}.}
Also, the pattern of field lines in $3D$ may change abruptly at a later time $t_0+\delta  t$ as these curves do not evolve smoothly in time; see the middle panel in Fig.(\ref{topology1}). Magnetic path-lines, i.e., trajectories of Alfv\'enic wave-packets, provide an alternative tool (see below).
If astrophysical magnetic fields can indeed undergo sudden changes in real space, plasma outflows can be considered as a secondary effect, observed as reconnection events. Alfv\'en wave-packets follow the local field, hence their trajectories provide a footprint of these changes in the field. Mathematically, the study of these trajectories, i.e., solutions of $\dot{\bf x}(t)={\bf V}_A({\bf x}(t),t)$, is analogous to the Lagrangian dynamics in hydrodynamics. For example, we are interested in the  separation of two such trajectories, i.e., $|{\bf x}(t)-{\bf y}(t)|$ at time $t$, which is related to Lyapunov exponents of the dynamical system $\dot{\bf x}={\bf V}_A$, or assuming incompressibility and absorbing density to magnetic field's definition, $\dot{\bf x}(t)={\bf B(x}(t),t)$ where magnetic field $\bf B$ satisfies the induction equation.\footnote{Throughout this paper, we assume a suitable non-dimensionalization, e.g., using an integral length scale $L$ and large-scale field $B_0$.} 
 The corresponding phase space ${\bf (x,B)}$ contains all possible states of this dynamical system {\color{black}which describes the motion of a single wave-packet. For $N$ wave-packets, we deal with an $N$-body system with $6N$ dimensional phase space and thus in the fluid approximation, the phase space would be infinite-dimensional.}

We will employ a physically intuitive and mathematically careful approach to magnetic topology in this paper. Topology is concerned with those properties of spaces that remain invariant under any continuous deformation, i.e, stretching and bending without cutting. 
Two objects (spaces) A and B (e.g., a solid ball and a cube) have the same topology if nearby points on A are mapped onto nearby points on B and vise versa: i.e., nearby points do not discontinuously mapped to points far away from each other. This means that there is a continuous, one-to-one and onto map from A to B with a continuous inverse (i.e., a homeomorphism). Thus if we deform object A (magnetic field at time $t$) to make object B (magnetic field at time $t\pm\delta t$), topology is preserved if nearby points are mapped onto nearby points, i.e., the distance between points is continuous in time. If the map from A to B is not onto and one-to-one, then points are "destroyed" during deformation thus topology will change. It means that the time evolution of A (magnetic field at time $t$) to B (magnetic field at time $t\pm \delta t$) does not respect time reversal symmetry. For example, dissipation in a magnetized plasma annihilates wave-packets and breaks time-reversal symmetry, thus changing magnetic topology.

Roughly speaking, for a magnetic field continuous in space and time, we expect "smooth deformation" in time---no discontinuous jump in its values or abrupt change in its direction. Therefore, wave-packets should move with a continuous velocity ${\bf V}_A({\bf x}, t)$  on continuous trajectories\footnote{Mathematically, this is because Alfv\'enic trajectories ${\bf x}(t)$ solve $\dot{\bf x}(t)={\bf B}$ with smooth $\bf B$.}, otherwise, we would expect a "topology-change". 
\begin{figure}[t]
 \begin{centering}
\includegraphics[scale=.55]{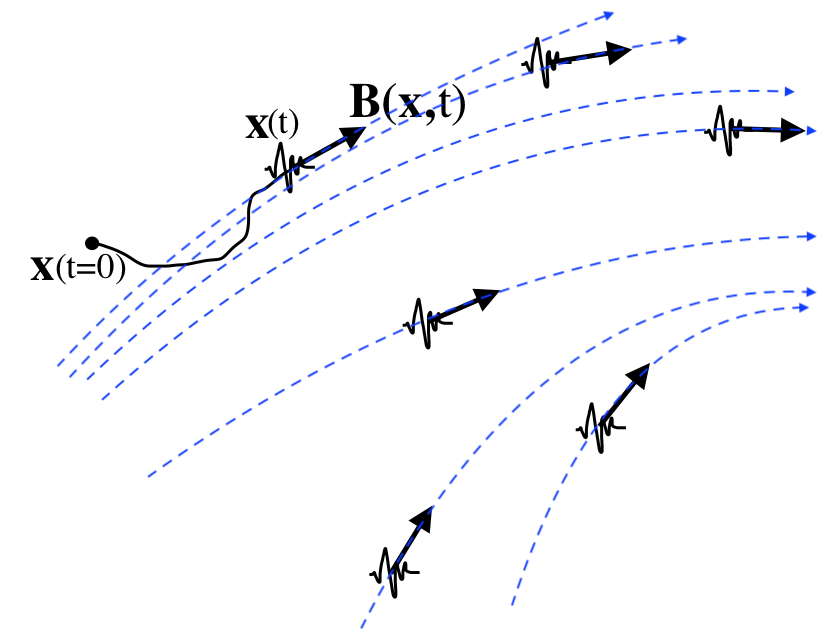}
\caption {\footnotesize {Magnetic "pattern", at any given time, is determined by providing a vector ${\bf B}$ at each point $\bf x$ in real space. Because the field evolves in time, the pattern changes, so we can follow wave-packets at points $\bf x$ in space moving with the local Alfv\'en velocity ${\bf V}_A\propto\bf B$. These pairs of coordinates correspond to points in the space $\bf (x,B)$ which is the phase space of the dynamical system $\dot{\bf x}=\bf B$, with $\bf B$ solving the induction equation. Thus magnetic topology can be understood as the metric topology of this phase space. "Metric" topology means that we simply use the Euclidean distance between points, such as $(\bf x,B(x))$ and $(\bf y,B(y))$; see eq.(\ref{metric}). This metric enforces the intuitive notion that nearby wave-packets at time $t$ remain nearby at a slightly earlier or later time $t\pm \delta  t$.}}\label{phasespace}
\end{centering}
\end{figure}
Hence, for magnetic topology to be preserved in time, we expect that nearby wave-packets, at time $t$, should remain nearby at a slightly different time $t\pm \delta t$. In fact, because of Lipschitz continuity of magnetic field (see \S\ref{SLagrangian}), this intuitive notion is equivalent to the following statement:
nearby wave-packets moving with almost the same velocity, at time $t$, should remain nearby, moving with almost the same velocity, at a slightly different time $t\pm \delta t$\footnote{This argument can be easily made mathematically precise using the $\epsilon\delta$ definition.} This simply means that the metric topology of the phase space $\bf (x,B)$, at any time $t$, is the same as its topology at a slightly different time $t\pm \delta t$, i.e, the topology is preserved (see \S\ref{Stopology}). {\color{black}As we will see, dissipation causes the volume of the phase space to contract. Eventually, over a time $\tau_T$, magnetic field completely diffuses away, and the phase-space dimension (initially infinite) approaches zero. Thus the dissipation rate $1/\tau_T$ can be taken as the rate of change of dimension, from infinity to zero. On the other hand, dimension is a topological invariant, i.e., its change means topology-change, thus the rate of topology change can be similarly defined as $1/\tau_T$. We will see that reconnection corresponds to rapid but continuous separation of these trajectories. Discontinuous divergence, which is not allowed if $\bf B$ is uniformly Lipschitz continuous, would lead to topology-change (nearby point not mapped onto nearby points). Dissipation, which breaks time-reversal invariance by annihilating wave-packets, can cause topology change in real dissipative systems; see also Fig.(\ref{topology1}). Mathematically, dissipation destroys the onto and one-to-one properties of the mapping, i.e., time evolution. The aim of this paper is to make these statements mathematically precise, and show how the emerging formalism can help gain a deeper and simpler picture for both reconnection and topology-change. Incidentally, from now on, we will assume incompressibility, absorbing density to the redefinition of magnetic field, with an appropriate non-dimensionalization as ${\bf B}={\bf V}_A$; see also Fig.(\ref{phasespace}).}

As for the detailed plan of the present work, in the following section, we first consider the divergence of Alfv\'enic wave-packets, recovering the fast Lazarian-Vishniac reconnection rate \citep{LV99,Lazarianetal2015,Review2020} in turbulent plasmas. Then, we will focus on magnetic topology, using a dynamical systems theory, and estimate its rate of change in both laminar and turbulent flows. The physical implications and connections to previous work will be discussed in Discussion.

\section{Lagrangian Formalism}\label{SLagrangian}
Astrophysical reconnection is understood to involve rapid changes in magnetic field configuration, thus trajectories of Alfv\'enic wave-packets, i.e., path lines, can rapidly diverge during reconnection. The dynamics of the  solutions, path-lines or Alfv\'enic trajectories, is in fact similar to Lagrangian dynamics in fluid mechanics---Alfv\'en velocity plays the role of the velocity field. As far as reconnection is concerned, one important quantity is what is known as $2$-particle diffusion in fluid dynamics, i.e., the separation (divergence) of any pair of trajectories over time; see e.g., \citep{JVV2019}. In the following, we will see how this simple notion, when applied to wave-packets, explains fast reconnection in turbulence and also clarifies the distinction between topology-change and reconnection.

The Lipschitz continuity of magnetic field means that $|{\bf B}({\bf x}, t)-{\bf B}({\bf y}, t)|\leq C | {\bf x} -{\bf y}|^h$ for some real $C\geq 0$ and $0<h\leq 1$. Consider the spatial separation of two arbitrary wave-packets ${\bf x}(t)$ and ${\bf y}(t)$ at time $t$, which were initially separated by $\Delta(t=0):=\Delta_0$.\footnote{Several notations exist in Lagrangian dynamics to denote the Lagrangian flow map, i.e, the map from particle's initial point ${\bf x}_0:={\bf x}(t=0)$ to   its final point ${\bf x}_{t_0}^{t}({\bf x}_0)$ at time $t$. We will simply use ${\bf x}(t)$ for the position of the particle at time $t$, implicitly assuming  that its initial position ${\bf x}(t=0)={\bf x}_0$ is given.} Using this and taking the time derivative of $\Delta(t)$, we arrive at
$d\Delta (t)/ dt\leq C[\Delta(t)]^h$, with simple solution
\begin{equation}
\Delta(t)\leq\Big[ \Delta_0^{1-h}+C(1-h)(t-t_0)  \Big]^{1\over 1-h}.
\end{equation}
In non-turbulent flows, $h\rightarrow 1$, thus 
$\Delta(t)\leq \Delta_0 e^{ C(t-t_0)  }$. At long times, assuming a near H{\"o}lder equality, we get $\Delta(t)\simeq \Delta_0e^{C(t-t_0)}$ for a chaotic flow with Lyapunov exponent $C$; implying that the initial conditions (i.e., the value of $\Delta_0$) are never forgotten. The important point is that in the limit, when the initial separation goes to zero, even for a chaotic flow, the final separation vanishes:
\begin{equation}\label{chaos}
\lim_{\Delta_0\rightarrow 0} |{\bf x}(t)-{\bf y}(t) |\rightarrow 0, \;\;\text{(laminar/chaotic flow)}.
\end{equation}

On the other hand, for $0<h<1$, we find $\Delta(t)\simeq \Big[C(1-h)(t-t_0) \Big]^{1\over 1-h}$, which implies that the information about initial conditions is lost! In other words, no matter how small the initial separation is, the wave-packets separate super-linearly with time:
\begin{equation}\label{Richardson1}|{\bf x}(t)-{\bf y}(t) |\sim t^{1\over 1-h},\;\;\text{(turbulent flow}).
\end{equation}
For $h=1/3$, corresponding to the Kolmogorov scaling for velocity field \citep{Kolmogorov1941}, we get Richardson law \citep{Richardson1926} {\color{black} (which predicts super-diffusion of particles $|{\bf x}(t)-{\bf y}(t) |\simeq \epsilon t^3$ with energy dissipation rate per mass $\epsilon$.)}.
The H{\"o}lder continuity for the magnetic field, i.e., $0<h<1$, results from the well-known effect of anomalous dissipation in turbulent plasmas; see e.g., \citep{Eyink2018}. In a turbulent flow with velocity field $\bf u$, the kinetic energy is viscously dissipated at a rate $\epsilon_{\nu}\equiv \nu|\nabla{\bf u}|^2$ while the magnetic energy is dissipated at a rate $\epsilon_\eta\equiv \eta|\nabla{\bf B}|^2$ (with viscosity $\nu$ and resistivity $\eta$). In fully developed turbulence, the Reynolds number $R_e\equiv LU/\nu$ and magnetic Reynolds number $R_m\equiv LU/\eta$, with characteristic length and velocity $L, U$, are high, i.e., one can take the limit of vanishing viscosity and resistivity; see also eqs.(\ref{nondimvel}) \& (\ref{nondimmag}) in \S\ref{subtop}. Thus one may naively expect that the viscous or resistive energy dissipation rate should vanish in these limits, i.e., $\lim_{\nu\rightarrow 0}\nu|\nabla{\bf u}|^2\rightarrow 0$ and $\lim_{\eta\rightarrow 0}\eta|\nabla{\bf B}|^2\rightarrow 0$. However, in turbulence, experiments and numerical simulations indicate otherwise (see \cite{JV2019} and references therein). In a fully turbulent fluid, no matter how small we take the viscosity or resistivity, the viscous dissipation rate as well as magnetic dissipation rate remain finite: $\lim_{\nu\rightarrow 0}\nu|\nabla{\bf u}|^2\nrightarrow 0$ and $\lim_{\eta\rightarrow 0}\eta|\nabla{\bf B}|^2\nrightarrow 0$. These {\it{dissipative anomalies}} indicate that the spatial derivatives of velocity and magnetic fields should blow up, i.e., the field become  H{\"o}lder singular\footnote{The real vector field ${\bf B}({\bf x})$ is H{\"o}lder continuous in ${{\bf x}}\in{\mathbb{R}}^n$ if $|{\bf B}(x) - {\bf B}(y) | \leq C| x - y|^h$ for some $C>0$ and $h>0$. If $h=1$, for any $x, y$, ${\bf B}$ is uniformly Lipschitz continuous. Also ${\bf B}$ is called H{\"o}lder singular if $0<h< 1$. A uniformly Lipschitz function $f(x)$, i.e., one which satisfies $|f(x)-f(y)|\leq C_f |x-y|^{h_f}$ for some $C_f>0$ with $h_f=1$, has a bounded derivative, i.e., $|f'(x)|< M$ for some $M>0$. In contrast, the derivative of a H{\"o}lder singular function $f$, i.e., one which satisfies $|f(x)-f(y)|\leq C_f |x-y|^{h_f}$ for some $C_f>0$ with $0<h_f<1$, can blow up; $|f'(x)|>\infty$.} which implies ill-defined spatial derivatives and hence ill-defined MHD equations (see e.g., \cite{EyinkS2006, Eyinketal2013, Eyink2018, JV2019, dynamicsJV2019}). In order to remove such singularities, one may use the coarse-grained field ${\bf B}_l$, defined by eq.(\ref{coarsegrain1}), and work with the "average" field at larger scales; see \S\ref{SRenormalized}.

Eq.(\ref{Richardson1}) implies that the mean square separation of wave-packets scales as $\langle |\Delta(t)|^2\rangle\sim t^{2\over 1-h}$. Thus averaging over, in a reconnection zone of scale $L$, it will take one Alfv\'en time, $t_A=L/V_A$, for wave-packets to leave the reconnection zone, during which they will on average separate by the distance
$$\Delta(t_A)\sim t_A^{1/1-h}\sim
\Big({L\over V_A}\Big)^{1/1-h}.$$
The reconnection speed $V_{rec}\sim {V_A \Delta/L}$ (which results from mass conservation in a zone of length $L$ and width $\Delta$) is thus given by
\begin{equation}\label{RecSpeed}
V_{rec}\sim {V_A \Delta/L}\sim \Big({L\over V_A}\Big)^{h/1-h}.\end{equation}
Assuming $h=1/3$, similar to the turbulent velocity field (see e.g., \cite{Eyink2018}, Sec.IV), we arrive at the Lazarian-Vishniac reconnection speed \citep{LV99} (see also \cite{Lazarianetal2015} Sec.3). Contrast the above estimate\footnote{{\color{black}In a more quantitative approach, one can use the energy dissipation rate, $\epsilon=u_L^4/V_A L_i$, from MHD turbulence theory in $\Delta^2(t)=\epsilon t^{3}$ (h=1/3), to  obtain $V_{rec}\simeq V_A\sqrt{L/L_i}M_A^2$ where $M_A$ is the Mach number, $L_i$ is the energy injection scale and $u_L$ is the (isotropic) injection velocity. This is Eq.(3.12) in \citep{Lazarianetal2015} which was obtained using an argument based on "field line diffusion".}}  with reconnection rate due to Ohmic diffusion of wave-packets, $\langle |\Delta(t_A)|^2\rangle\sim \eta t_A$, i.e., 

$$V_{rec}^{SP}\sim V_A\Delta/L\sim (\eta V_A/L)^{1/2}={V_A\over\sqrt{S}},$$
where $S:=V_AL/\eta$ is the Lundquist number. This result, the Sweet-Parker speed \citep{Parker1957,Sweet1958}, is extremely slow in any astrophysical setting due to typically very large Lundquist numbers; see e.g., \cite{Review2019,Review2020}. These results, of course, are not new: we have simply recovered them here without appealing to the notion of field line diffusion. In the following subsections, we will also see that our approach not only clarifies the concept of magnetic topology and distinguishes it from reconnection, but it also provides an estimate for the rate of topology-change. In fact, it turns out that magnetic topology in laminar flows changes with a rate proportional to resistivity; see eq.(\ref{laminarrate}).

Rapidly diverging trajectories even in the limit of their vanishing initial  separation, associated to spontaneous stochasticity or the "real" butterfly effect \citep{Bernardetal.1998}, differs from simple deterministic chaos (the butterfly effect). In the former case, the dynamics is singular (e.g., magnetic field is H{\"o}lder so $\dot{\bf x}={\bf B}$ has non-unique solutions) and randomness of trajectories persists at finite times even in the limit of vanishing noise.\footnote{{\color{black}Of course, the limit of vanishing viscosity $\nu$ only means considering smaller and smaller viscosities, rather than taking a zero viscosity $\nu=0$ which is unphysical. This definition of mathematical limit in physics is assumed to be understood in other areas, e.g., thermodynamics, too.}} This is distinct from simple deterministic chaos in which predictability times can be lengthened arbitrarily by reducing noise since final solution always remains proportional to the initial conditions.  This is, incidentally, the primary reason why the weather (governed by Navier-Stokes equations) cannot be forecasted for more than almost two weeks: this time cannot be lengthened even with more advanced technology (the real butterfly effect).\footnote{A nice discussion of this effect in weather forecasting, along with historical notes regarding E. Lorenz pioneering work, can be found in the popular book by T. Palmer: The Primacy of Doubt \citep{Palmer}.}On the other hand, in simple chaos (the butterfly effect), such predictability times can be lengthened by reducing noise (e.g., by making initial separation of trajectories arbitrarily smaller). 
Spontaneous stochasticity can be regarded as “super-chaos” associated with the formation 
of singularities in the dynamics and consequent divergence of Lyapunov exponents to infinity. Vanishingly small random perturbations can then be propagated to large scales 
in a \textit{finite} amount of time. Extremely small but unavoidable sources of noise such as thermal fluctuations have been shown to easily trigger spontaneous stochasticity in turbulence \citep{Bandaketal2024}. These statements explain why, in the above calculations for a turbulent flow, we employed mean square separation: the very notion of a Lagrangian trajectory breaks down in turbulence since the system is spontaneously stochastic and trajectories remain intrinsically random in fully developed turbulence. This effect is in fact intimately related to the old, well-known notion of Richardson (2-particle) diffusion (for a brief and nice discussion of Richardson diffusion and its connection to spontaneous stochasticity, see \cite{Eyink2011}). At large (inertial) scales, the singularity can be removed by a coarse-graining procedure which leads to a smooth, large-scale magnetic field on any inertial scale $l$; see \S\ref{SRenormalized}.

\subsection{Magnetic Topology\footnote{A topology on a set $X$ is defined as the collection of subsets of $X$, denoted by $T$, such that (i) any arbitrary union of elements of $T$ belongs to $T$, (ii) 
any finite intersection of elements of $T$ belongs to $T$, (iii) the empty set and $X$ both belong to $T$. We call $(X, T)$ a topological space. The elements of $T$ are called open sets. For instance, take $X={\mathbb R}^3$ and define the open sets, i.e., elements of $T$, as open balls in ${\mathbb R}^3$. (An open ball around a point ${\bf x}_0$ is the set of all points $\bf x$ such that $|{\bf x}-{\bf x}_0|<r$, i.e., all points with a distance less than $r$ from ${\bf x}_0$. In ${\bf R}$ open balls are open intervals.) This topology, defined invoking the notion of distance between points in a set, is called metric topology, to be extensively used in the present paper.}}\label{Stopology}
Because magnetic field evolves in time, interacting with the velocity field, we have a non-autonomous dynamical system:
\begin{equation}\label{baredynamics}
\begin{cases}
 \dot {\bf x}:={d{\bf x}(t)\over dt}={\bf B(x}(t), t),\\ \dot{\bf B}:={\partial {\bf B(x}(t),t)\over \partial t}=-\nabla\times {\bf E},
 \end{cases}
\end{equation}
where ${\bf x}(t)$ denotes a trajectory. The corresponding phase space is $(\bf x,B)$, which has a metric topology imposed by the Euclidean metric
\begin{equation}\label{metric}
\Delta_t:=\sqrt{|{\bf B}({\bf x}, t)-{\bf B}({\bf y}, t)|^2+| {\bf x} -{\bf y}|^2},
\end{equation}
which defines magnetic topology in an intuitive way, as discussed in the Introduction. Because homeomorphisms have, by definition, a continuous inverse, hence one condition for magnetic topology to be preserved is that its time evolution, hence equations of motion, should be invariant under time reversal; $t\rightarrow -t$. In vacuum, electric field is even under time reversal, ${\bf E}(-t)=+{\bf E}(t)$ while magnetic field is odd, i.e., ${\bf B}(-t)=-{\bf B}(t)$. Consequently, the Faraday equation, $\partial_t {\bf B}=-\nabla\times {\bf E}$, respects time reversal invariance as expected. In real fluids with finite resistivity and viscosity, however, the time reversal symmetry is broken. Non-ideal terms such as resistive electric field $\eta\nabla\times \bf B$ in the Ohm's law break the time symmetry in the induction equation which governs the evolution of magnetic field:
\begin{equation}\label{bareinduction}
\partial_t{\bf B}=\nabla\times ({\bf u\times B}-\eta\nabla\times{\bf B}),
\end{equation}
where $\bf u$ is the velocity field solving the Navier-Stokes equation. Hence, in general, ${\bf B}(-t)\neq \pm {\bf B}(t)$. Consequently, the topology of the bare magnetic field ${\bf B(x}, t)$ is not preserved in a real  fluid. 

 In laminar flows, magnetic field is continuous in time and Lipschitz continuous in space, hence $\Delta_t$ is continuous and trajectories ${\bf x}(t)$ are uniquely defined. However, the time-reversal symmetry is broken by non-ideal effects such as resistivity. In the phase space $(\bf x,B)$, dissipation contracts any set of initial conditions to a single dimensionless point thus the topology changes because dimension is a topological invariant. However,  discontinuous changes in topology are not allowed due to continuity and magnetic topology is nearly preserved, i.e., it changes only with a rate proportional to resistivity. Thus fast laminar reconnection, if it exists, cannot involve topology change. 
 
 In real astrophysical flows, on the other hand, $\bf B$ is H{\"o}lder continuous, i.e., $0<h<1$, due to turbulence. As a result, trajectories ${\bf x}(t)$ are non-unique with diverging Lyapunov exponents, i.e, the system is spontaneously stochastic. However, what can physically be measured is the renormalized (coarse-grained) magnetic field ${\bf B}_l$ obtained by integrating out the small degrees of freedom over a spatial region of scale $l>0$ rather than the mathematical field ${\bf B(x}, t)$ at a single spacetime point $({\bf x}, t)$. The renormalized field ${\bf B}_l$ is smooth and solutions of $\dot{\bf x}(t)={\bf B}_l({\bf x}(t),t)$ are unique but the \textit{renormalized} (large-scale) topology associated with ${\bf B}_l$ can still change by non-linear turbulent effects due to time reversal symmetry breaking. In the following two subsections, we estimate the rate of topology change in laminar and turbulent flows. We will see that in laminar flows, magnetic topology changes with a rate proportional to resistivity, thus it is almost preserved for highly conducting plasmas. Nevertheless, in turbulence, topology changes on any inertial scale $l$ with a rate independent of small-scale plasma effects (\S\ref{subtop}).

\subsection{Topology Change in Laminar Flows}

Because dimension of a mathematical space is a topological invariant, i.e., it is preserved under homeomorphisms, its change implies topology change. For instance, in mapping a 3-dimensional solid ball to a 2-dimensional surface, topology changes because of the change in dimension (thus the mapping is not a homeomorphism). Invoking this simple mathematical notion, we will obtain the rate of magnetic topology change in laminar (see below) as well as in turbulent flows (\S\ref{subtop}). Let us consider the rate of topology change for the dynamical system corresponding to the \textit{bare} (i.e., not renormalized; see eq.(\ref{induction1}) below) induction equation given by eq.(\ref{bareinduction}):
\begin{equation}
\begin{cases}
\dot{\bf x}={\bf B(x}(t), t),\\ 
\dot {\bf B}=\nabla\times ({\bf u\times B}-\eta\nabla\times{\bf B}),
 \end{cases}
\end{equation}
which may be also written in a more compact form as 
\begin{equation}\label{Dsystem}
{\partial\over\partial t}\begin{pmatrix}
{\bf x} \\
{\bf B}
\end{pmatrix}=
\begin{pmatrix}
{\bf B}({\bf x}(t), t)\\
{\bf G}[{\bf B}({\bf x}(t), t); {\bf x}]
\end{pmatrix}:={\bf F}[{\bf B};{\bf x}],
\end{equation}
where $\bf F$ and ${\bf G}:=\nabla\times ({\bf u\times B}-\eta\nabla\times{\bf B})$ are functionals of $\bf B$ (and $\bf u$, suppressed here for brevity). For a general dissipative dynamical system {\color{black} with $N$ particles described by} $\dot{\bf x}(t)={\bf f(x}(t))$, a solid ball of initial conditions in the {\color{black} $6N$ dimensional} phase space $\bf (x,f)$ contracts with the rate $\tau^{-1}=|\nabla.\bf f|$.\footnote{To see this, simply note that any volume $V(t)$ of points in the phase space $(\bf x,f)$, corresponding to the dynamical system $\dot{\bf x}(t)={\bf f}$, with normal vector $\bf n$ to its surface $S(t)$ changes with time as $V(t+dt)=V(t)+{\bf f.n}\;dSdt$ thus in the limit, $\dot V/V= \nabla.{\bf f}$.} Thus due to dissipation, the initial {\color{black}$6N$}-dimensional ball will contract to a dimensionless point, i.e., the dimension of the region changes as well. On the other hand, dimension is a topological invariant, thus its change indicates topology-change with the same rate $\tau^{-1}$.
{\color{black}In our fluid approximation, in which magnetic wave-packets are treated as parcels of fluid, the dimension of phase space is infinite, so the dissipation rate is the inverse of the time it takes for dimension to decrease from infinity to zero.} Thus our aim is to obtain the rate $\tau^{-1}_T:=\nabla.\bf F$, for the system (\ref{Dsystem}) using the gradient operator in the phase space, i.e., $\nabla:=\Big({\partial\over\partial{\bf x}},{\delta\over\delta {\bf B}}\Big)$. Because $\nabla_{\bf x}.{\bf B}=0$, the $\bf x$ derivatives vanish and the remaining functional derivative can be evaluated as follows:
\begin{eqnarray}\nonumber
{\delta G_k[{\bf B; x}]\over\delta B_h({\bf x}')}&=&
{\delta\over\delta B_h({\bf x'})}\Big[\Big( \nabla_{\bf x}\times({\bf u\times B})\Big)_k+\eta \triangle_{\bf x} B_k \Big]\\\nonumber
&=&
{\delta\over\delta B_h({\bf x'})}\Big[ \epsilon_{ijk}\epsilon_{lmj}\partial_i\Big( u_l B_m\Big)+\eta\triangle_{\bf x} B_k\Big]\\\nonumber
&=&
\Big[\epsilon_{ijk}\epsilon_{lkj}\partial_i\Big(u_l\delta_{mh}\delta_\Lambda^3({\bf x-x'}\Big) \\\nonumber
&&\;\;\;\;\;\;\;\;\;\;\;+\eta \;\delta_{kh}\;\triangle_{\bf x} \delta^3_\Lambda({\bf x-x'})\Big].
\end{eqnarray}
Therefore,
\begin{eqnarray}\nonumber
\tau^{-1}_T&:=&\sum_{k=1}^3\int\int\int d^3{\bf x}{\delta G_k[{\bf B; x}]\over\delta B_k({\bf x})}\\\nonumber
&=&\int\int\int d^3{\bf x}
\Big[-2\delta_{il} \partial_i(u_l\delta^3_\Lambda({\bf 0}))+3\eta\triangle_{\bf x}\delta^3_\Lambda({\bf 0}) \Big]\\\nonumber
&=&\Big[-2\nabla.({\bf u}\delta^3_\Lambda({\bf 0}))+3\eta(\triangle_{\bf x}\delta^3_\Lambda({\bf 0}))\Big]V \\\label{laminarrate}
&=&\eta\;(3V\triangle_{\bf x}\delta^3_\Lambda({\bf 0})),
\end{eqnarray}
where $\delta^3_\Lambda({\bf x})={1\over V}\sum_{|{\bf k}|<\Lambda}e^{i{\bf k.x}}$, hence $\triangle_{\bf x}\delta_\Lambda^3({\bf 0})=(1/V)\sum_{|{\bf k}|<\Lambda} (-k^2)=const.$, in real-space volume $V$. Note that in order to have a physically meaningful continuum limit, there is a high-wavenumber cut-off in the velocity and magnetic fields as well as in the spatial delta-function $\delta^3_\Lambda({\bf x})$\footnote{Such ultra-violet (UV) cut-offs (either in momentum or real space) are required because physical quantities and equations of motion lose their meaning at very small scales, e.g., scales below {\color{black}the mean-free-path (gas) or inter-particle distance (fluid)} which are still much larger than the Planck scale! Thus these are in fact \textit{effective} theories valid only on "larger" scales. Surprisingly, this  well-known notion in many fields such as statistical physics, high energy physics and quantum field theories, is not yet appreciated in some other fields.}.

Thus magnetic topology changes with a rate proportional to resistivity. In magnetized astrophysical environments, resistivity is typically very small thus magnetic topology is expected to change slowly. Equivalently, we could average the field (i.e., coarse-grain or integrate out the small degrees of freedom) over small scales and look at the dynamics at much larger scales where  plasma non-idealities such as resistivity are negligible\footnote{This basically serves as the definition of the inertial scales in turbulence.}. The non-ideal terms in the corresponding "coarse-grained" induction equation governing the large scale field ${\bf B}_l$ on scale $l$ will be negligible then; i.e., at large scales, we recover the "ideal" induction equation  (see \S \ref{SRenormalized} below), implying very slow topology change, in agreement with the above conclusion. In the next subsection, we will show that this is not the case in real astrophysical systems due to the presence of turbulence. In fact, at large scales, turbulent effects will dominate which can lead to fast dissipation and topology-change.

\subsection{Renormalized Topology}\label{SRenormalized}
The detailed magnetic field configuration, or magnetic pattern, e.g., on the surface of a distant star, depends on the resolution available to the observer: a low-resolution, terrestrial instrument will obviously detect different magnetic patterns compared with what a high-resolution instrument on a satellite closely orbiting the star would. In fact, no matter how great our resolution is or how close we are to the system, what we can measure as the magnetic field at point $\bf x$ is the average field ${\bf B}_l$ in a finite volume of size $l^3>0$ rather than the mathematical vector ${\bf B (x)}$ defined at point $\bf x$. The reason, as mentioned before, is that any instrument can perform a measurement only in a finite volume in space and cannot detect the field defined at a single dimensionless (mathematical) point. Magnetic field \textit{measurable} in any experiment is a coarse-grained field ${\bf B}_l$, which is essentially the average field over a length scale $l$. If the measured magnitude and direction depend on our resolution scale $l>0$, and we can only measure the physical field ${\bf B}_l$ as an average over a length-scale $l$ and not the mathematical bare field $\bf B$, what do we mean by the topology of the field $\bf B$? The crucial point is that  although ${\bf B}_l$ will differ from ${\bf B}_L$ for $l\neq L$, but on all (inertial) scales $l$ and $L$, both ${\bf B}_l$ and ${\bf B}_L$ are governed by exactly the \textit{same} dynamics. This is the heart of (Wilsonian) Renormalization Group (RG) theory.

The coarse-grained field can be defined using any rapidly decaying test function $\phi$ to coarse-grain a given field ${\bf{B}}({\bf{x}}, t)$ at a spatial scale $l>0$ by writing\footnote{{\color{black}Coarse-grainig is the common terminology in physics, also called mollifying in mathematical literature and low-pass filtering in engineering.}} 
\begin{equation}\label{coarsegrain1}
{\bf{B}}_l ({\bf{x}}, t)=\int_V \phi\left({{\bf{r}}\over l}\right).{\bf B}({\bf{x+r}}, t) {d^3r\over l^3},
\end{equation}
where $\phi({\bf{r}})=\phi(r)$ is a smooth and rapidly decaying (scalar) kernel\footnote{Without loss of generality, we also assume $\phi({\bf{r}})\geq 0$, $\lim_{|\bf r|\rightarrow \infty} \phi({\bf{r}})\rightarrow 0$, $\int_V d^3r \phi({\bf{r}})=1$, $\int_V d^3r \; {\bf{r}}\;\phi({\bf{r}})=0$, $\int_V d^3r |{\bf{r}}|^2 \;\phi({\bf{r}})= 1$ and $\phi({\bf{r}})=\phi(r)$ with $|{\bf{r}}|=r$. Mathematically, $\phi \in C_c^\infty ({\mathbb{R}})$; the space of infinitely-differentiable functions with compact support. A function $g$ is said to have a compact support (set of its arguments for which $g\neq 0$) if $g=0$ outside of a compact set (equivalent to closed and bounded sets in ${\mathbb{R}}^m$). As an example, one may work with $\phi(r)= \phi_0\exp{{-1\over 1-r^2}}$ for $|r|<1$ and $\phi=0$ for $|r|\geq 1$. The normalization constant $\phi_0$ is about $0.88$ in three dimensions. For a quick, but more detailed introduction, see \cite{Eyink1996}, Sec. 2.1}. In fact, the renormalized field ${\bf B}_l$ is the average magnetic field of a fluid parcel with length scale $l$. The coarse-grained induction equation (obtained by multiplying the bare induction equation by $\phi({\bf r}/l)$ and integrating) reads\footnote{Mathematically inclined reader would notice that this method is equivalent to the weak formulation; see e.g., \citep{Eyink2015}.}
\begin{equation}\label{induction1}
{\partial {\bf{B}}_l\over \partial t}=\nabla\times ( {\bf{u}}_l \times {\bf{B}}_l - {\bf{R}}_l-{\bf P}_l),
\end{equation}
using the renormalized Ohm's law ${\bf E}_l+({\bf u\times B})_l={\bf P}_l$, which can also be written as 
\begin{equation}\label{Ohm}
{\bf{E}}_l={\bf P}_l+{\bf R}_l-{\bf{ u}}_l \times {\bf{B}}_l.
\end{equation}
 Even with a negligible non-ideal term ${\bf P}_l$, the non-linear term ${\bf{R}}_l=-( {\bf{ u \times B}})_l+{\bf{u}}_l \times {\bf{B}}_l$ will be generally large in turbulence. Furthermore, what is really important is its curl, $\nabla\times{\bf{R}}_l$, which can be large and dominant in the induction equation \citep{Eyink2015}. The turbulent electromotive force (EMF) ${\cal{E}}_l\equiv-{\bf{R}}_l$, is the motional electric field induced by turbulent eddies of scales smaller than $l$ and plays a crucial role in magnetic dynamo theories.
However, note that despite its similarity, this quantity differs from the mean EMF defined as a statistical average $\overline{ {\bf u}'\times {\bf b}'}$ with fluctuating velocity and magnetic fields ${\bf u}', {\bf b}'$, commonly used in mean field theories. This is because ${\bf R}_l$ is deterministic, unlike mean field EMF which is statistical. Also, in defining ${\bf R}_l$ no assumptions are made of scale separation between large-scale mean fields and small-scale fluctuations ${\bf u}', {\bf b}'$. In addition, ${\bf R}_l$ and coarse-grained equations above are effective equations which depend upon an arbitrary length scale $l$, which may be varied according to the desired resolution of the physics; see also \citep{EyinkAluie2006}.

At large scales, where small-scale dissipative effects can be neglected (i.e., ideal Ohm’s law holds in the turbulent inertial range in the coarse-grained or weak sense), we can write
\begin{equation}\label{induction100}
{\partial {\bf{B}}_l\over \partial t}=\nabla\times ( {\bf{u}}_l \times {\bf{B}}_l - {\bf{R}}_l).
\end{equation}
In the coarse-grained induction equation, eq.(\ref{induction100}), we can use the estimate  $|\nabla\times {\bf R}_l|\simeq {1\over l}|\delta {\bf u}(l)\times  \delta {\bf B}(l)|$ with increments across scale $l$ \citep{EyinkAluie2006, Eyink2015}. In a non-turbulent flow, $\delta {\bf u}(l)\sim l$ and $\delta {\bf B}(l)\sim l$ hence in the limit $l\rightarrow 0$, the non-linear term ${\bf{R}}_l$ would vanish:
\begin{equation}\label{induction101}
{\partial {\bf{B}}_l\over \partial t}\simeq \nabla\times ( {\bf{u}}_l \times {\bf{B}}_l).
\end{equation}
Therefore, we recover the "ideal" induction equation on scale $l$ which is assumed to be much larger than the dissipative scale. The implication of these familiar results in the context of magnetic topology is that in a laminar flow, at scales larger than the dissipative scale, the magnetic topology is preserved within a good approximation. This is, of course, expected since we use the ideal fluid approximation and magnetic diffusivity annihilates the field on a resistive time scale. In turbulence, on the other hand, we cannot ignore non-linear term ${\bf{R}}_l$ whose curl, which enters induction equation, remains large even in the limit of vanishing resistivity.

\subsection{Topology Change in Turbulent Flows}\label{subtop}

The Navier-Stokes equation can be easily cast into a non-dimensional form at large scales using the parameters (in standard notation) ${\overline {\bf x}}={\bf x}/L,\;\;\;{\overline t}=t/(L/U),\;\;\;{\overline{\bf u}}={\bf u}/U,\;\;\;{\overline p}=p/\rho U^2$ as
\begin{equation}\label{nondimvel}
{\partial \overline {\bf u}\over \partial \overline t}+\overline{\bf u}.\overline\nabla\overline{\bf u}=-\overline\nabla\overline p+{1\over Re}\overline\nabla^2\overline{\bf u}.
\end{equation}
Analogously, the induction equation is written as 
\begin{equation}\label{nondimmag}
{\partial \overline{\bf B}\over \partial \overline t}={1\over Re_m}{\overline \nabla}^2 \overline{\bf B}+{\overline\nabla}\times({\overline{\bf u}}\times\overline{\bf B}),\end{equation} where $\overline{\bf B}={\bf B}/B_0$ with  characteristic field $B_0$. Therefore, a small viscosity (resistivity) translates into a large (magnetic) Reynolds number. One might naively neglect the terms proportional to $1/Re$ (and $1/Re_m$), to recover the "ideal" equations. However, at large Reynolds numbers, the flow is extremely sensitive to small perturbations implying the presence or development of turbulence. Hence, the limit of vanishing viscosity (resistivity) may correspond to a (complicated) turbulent flow rather than the "simple" ideal case!

What is the rate of magnetic topology change due to turbulent effects? In the phase space $(\bf x, B)$, this is basically the rate at which the topology of a solid ball of initial conditions for the dissipative dynamical system 
\begin{equation}
\begin{cases}
\dot{\bf x}(t)={\bf B}_l({\bf x}(t), t)\\
\dot{\bf B}_l({\bf x}(t), t)=\nabla\times ( {\bf{u}}_l \times {\bf{B}}_l - {\bf{R}}_l-{\bf P}_l):={\bf G}_l[{\bf B}_l; {\bf x}, t],
\end{cases}
\end{equation}
with ${\bf G}[{\bf B}_l; {\bf x}, t]$ as a functional of ${\bf{B}}_l$, changes over time. The dissipation rate of this dynamical system, $\tau_T^{-1}=\nabla.{\bf G}$ with the phase-space gradient $\nabla:=\Big({\partial\over \partial{\bf x}(t)},{\delta \over \delta {\bf B}_l}\Big)$, is 
\begin{eqnarray}\nonumber
\tau^{-1}_T&:=&\Big|\sum_{k=1}^3\int\int\int d^3{\bf x}
{\delta G_l^k[{{\bf B}_l; {\bf x}}]\over\delta B_l^k({\bf x})}\Big|
\end{eqnarray}
where
\begin{eqnarray}\nonumber
{\delta G_l^k[{{\bf B}_l; {\bf x}}]\over\delta B_l^{k'}({\bf x}')}&=&{\delta\over\delta B_l^{k'}({\bf x}')}
\Big(\nabla_{\bf x}\times \Big(\Big[ {\bf{u}}_l({\bf x}) \times {\bf{B}}_l({\bf x})\Big] \\\nonumber
&&\;\;\;\;\;\;\;\;\;\;\;\;\;\;- {\bf{R}}_l({\bf x})-{\bf P}_l({\bf x})\Big)\Big)^k\\\nonumber
&=&{\delta\over\delta B_l^{k'}({\bf x}')}
\Big(\nabla_{\bf x}\times \Big( - {\bf{R}}_l({\bf x})-{\bf P}_l({\bf x})\Big)\Big)^k.
\end{eqnarray}
Note that the contribution of small-scale plasma effects, i.e., different processes collectively denoted as ${\bf P}_l$ in our notation, will be negligible as e.g., for the case of Ohmic electric field considered before. Thus we arrive at the estimate
\begin{equation}\label{rate}
\tau^{-1}_T\doteq \Big|\sum_{k=1}^3\int\int\int d^3{\bf x}
{\delta\over\delta B_l^{k'}({\bf x}')}
\Big(\nabla_{\bf x}\times {\bf{R}}_l({\bf x})\Big)^k\Big|.\end{equation}
{\color{black}This expression gives the turbulent dissipation rate of magnetic topology at any inertial scale $l$. We emphasize that fully developed turbulence does not respect time reversal symmetry thus the time evolution as a mapping is not a homeomorphism, implying topology change by turbulent (non-viscous) dissipation. This dissipation is due to the energy cascade and magnetic-to-kinetic energy conversion, hence it has nothing to do with viscous dissipation in the dissipative range.}

We will not evaluate the expression given by eq.(\ref{rate}) further, since obtaining explicit results is complicated and not required for our purposes here. The important point is that the rate given by eq.(\ref{rate}) is obviously independent of small-scale plasma effects and depends totally on turbulent effects. Therefore unlike laminar flows, where topology changes with a rate proportional to resistivity which is typically negligible in astrophysical systems, in turbulence magnetic topology may change on much faster time scales.

As a side note, to close this section, let us mention that to evaluate the functional derivative given by eq.(\ref{rate}) is mathematically challenging. A similar, but simpler, problem is encountered in considering the velocity field itself, i.e., in the coarse-grained Navier-Stokes equation which involves taking the functional derivative of the term ${\delta\over\delta {\bf u}_l}\nabla. \Big(({\bf{u} \bf{u}})_l-{\bf u}_l{\bf u}_l\Big)$. Such a calculation involves mode-reduction, e.g. using Zwanzig-Mori projection methods or a path-integral approach to integrate out unresolved scales. The result is spatially nonlocal, non-Markovian,
transcendentally nonlinear and also random with colored, multiplicative noise \citep{Eyink1996}. For the velocity field, the divergence of the systematic part of $\nabla.(\bf {u u})_l$ equals zero and the non-vanishing contribution arises entirely from the “eddy noise”.

\section{Discussion}
 In this paper, we have argued that reconnection is associated with continuous divergence of Alfv\'enic wave-packets (magnetic path-lines) over time, i.e., roughly speaking, rapid but smooth change in magnetic pattern over time. Topology-change is associated with discontinuous divergence of trajectories or their annihilation by dissipative effects; resistivity in laminar flows and turbulent dissipation in turbulence. 
 
{\color{black}In laminar flows, if one accepts that large-scale reconnection is slow and Sweet-Parker type, then it can be thought of as a topology-change with a slow resistivity-dependent rate given by eq.(10). If one accepts the view that laminar reconnection can be fast (i.e., proceeding on time scales much shorter than the resistive time scale), then reconnection cannot be equivalent to topology-change, which is always slow in laminar and even chaotic flows, as shown in this paper, because of the continuity of magnetic field. This is theory, what about observations?

It is an observational fact that reconnection is fast in real plasmas, especially in astrophysics (e.g., in the solar wind and solar corona). Solar observations indicate that the reconnection rates can vary significantly, implying that reconnection depends on local physics. Non-thermal broadening of spectral lines (and other measures, see e.g., \cite{Burkhart2010}) indicate that astrophysical flows are turbulent \citep{Review2020}. Astrophysical plasmas generally have very large Reynolds (and magnetic Reynolds) numbers thus turbulence is expected to be present. Even if initially absent, onset of reconnection e.g., due to plasma instabilities, will bring turbulence to the play. Therefore, both observations and theory indicate that reconnection proceeds in turbulence. Turbulence dominates non-ideal plasma effects on reconnection, from Hall effect to tearing modes instabilities, as discussed in the argument after eqs. (12) \& (13). In fact, turbulence leads to super-chaotic divergence of Alfv\'enic wave-packets with an infinite Lyapunov exponent, thus implying a much faster reconnection than what deterministic chaos can drive, as eqs. (2)-(4) indicate.
Even in chaotic flows, where the Alfv\'enic trajectories diverge exponentially, their separation at later times remains proportional to the initial separation, unlike turbulent flows.}

In turbulence, the picture is more complicated than laminar flows and both reconnection and topology-change is driven by spontaneous stochasticity or super-chaos---persistent random behavior of non-unique Alfv\'enic trajectories independent of their initial separation. Specifically, in turbulence, the non-linear dynamics governing magnetic field becomes singular and randomness in wave-packet trajectories survives even in the limit of vanishing noise, i.e., fast separation of trajectories at later times even when the initial separation of trajectories tends to zero. Fast turbulent reconnection results from enhanced turbulent mixing of trajectories while topology-change results from turbulence's time-reversal symmetry breaking. These effects are deeply related to spontaneous stochasticity.

The simple approach presented in this paper  provides a clear and intuitive topology for magnetic fields independent of the notion of magnetic field lines and their complicated motion through plasma.   In fact, versatility of a powerful tool, such as Feynman's diagrams; concept of a gravitational field or the notion of magnetic field lines, can make it look more physical than it is in reality leading to misuse. Feynman used to interpret his diagrams in a more physical way than what is understood today. Gravitational field around a mass became only part of the metric in general relativity. Similarly, the powerful notion of magnetic field lines, since their introduction by Faraday, can be replaced with other more appropriate tools such as magnetic path-lines for gaining a deeper and simpler picture. In a real magnetized plasma in three dimensions, for example, magnetic field lines might become quite inadequate to describe processes such as reconnection or magnetic topology change. Our approach in this paper provides an alternative way of looking at reconnection and magnetic topology-change, in terms of magnetic path-lines ${\bf x}(t)$ solving $\dot{\bf x}(t)={\bf B(x}(t),t)$ instead of field lines, which are parametric curves depicting magnetic field pattern \textit{only} at a given time. 

Alfv\'enic trajectories or magnetic path-lines are associated with a dynamical system whose phase space has a natural metric topology for magnetic field. This magnetic topology evolves with time and it is easy to see what conditions should be satisfied for the topology to be preserved. In laminar flows, magnetic topology can change with a rate proportional to resistivity, hence if reconnection is fast, i.e., it occurs on time scales much shorter than resistive time, then reconnection and topology change should be distinct phenomena. In turbulence, time-reversal symmetry is broken and topology cannot be preserved. In fact, magnetic topology in a turbulent plasma can change by enhanced turbulent diffusion on any inertial scale. 

Our simple calculations also suggest connections to other approaches to magnetic reconnection. For example, one can use the coarse-grained induction equation, eq.(\ref{induction1}), to study the time evolution of the unit tangent vector, $\hat{\bf B}_l={ {\bf B}_l / B_l }$. In fact, the coarse-grained induction equation implies
\begin{equation}\label{zap10}
  \partial_t \hat {\bf{B}}_l={\nabla\times({\bf u}_l\times{\bf B}_l)^\perp \over B_l}-({{\bf{\Sigma}}_l}^\perp+{{ {\sigma}}_l}^\perp),
 \end{equation}
where $(.)^\perp$ indicates the perpendicular direction with respect to the large-scale field ${\bf B}_l$ and 
\begin{equation}\label{slipV}
\begin{cases}
{\boldsymbol{\Sigma}}_l={(\nabla\times{\bf{R}}_l) \over B_l},\\
{ \boldsymbol{\sigma}}_l={(\nabla\times{\bf{P}}_l) \over B_l}.
\end{cases}
\end{equation}
The term ${\boldsymbol{\Sigma}}_l$ (${ \boldsymbol{\sigma}}_l$) has been shown to govern magnetic reconnection in turbulence (laminar flows) \citep{Eyink2015}.
On the other hand, we have shown in this paper, eq.(\ref{rate}), that the rate of magnetic topology change depends on the functional derivative of $B_l {\boldsymbol{\Sigma}}_l$ (and $B_l{\boldsymbol{\sigma}}_l$ on small scales in laminar flows):
\begin{eqnarray}\nonumber
\tau^{-1}_T&\doteq&\Big|\sum_{k=1}^3\int\int\int d^3{\bf x}
{\delta \Big(B_l{\Sigma}_l^k\Big)\over\delta B_l^{k}({\bf x})}
\Big|.
\end{eqnarray}
This estimate for turbulent flows implies that magnetic topology changes on all turbulent inertial scales $l$ and unlike laminar flows, it is independent of small scale (plasma) effects. These are the distinguishing characteristics of turbulent (stochastic) reconnection as well \citep{LV99,Eyink2015,Review2020}. In fact,  "naive" dimensional analysis of the above estimate suggests a fast rate of order $\tau_T^{-1}\sim \Delta u_l/l$, i.e., eddy turn-over rate on scale $l$. In laminar flows, this rate is proportional to (typically negligible) resistivity; eq.(\ref{laminarrate}). Spontaneous stochasticity seems to enhance both topology-change and reconnection in turbulence.

Finally, our approach also emphasizes the crucial roles scale and turbulence play in the study of magnetic topology change as well as its connection to turbulent reconnection. 
Any physical measurement can be performed only in a finite region of space (and during a finite interval in time). This somehow resembles the uncertainty principle in the sense that we can never measure the "real" mathematical magnetic field ${\bf B(x},t)$ no matter how technologically advanced our measuring instruments become over time. In fact, although electromagnetic fields are in most applications assumed to be well-defined at small scales, this is not true for general physical fields. In condensed matter physics, for example, it is common for a general field theory to have a natural cut-off at small length scales (or high energies), e.g., the spacing between atoms in a lattice crystal. In most quantum field theories, there exist (ultra-violet) infinities and the field is not well-defined at very small scales. Hence, a cut-off is usually introduced to regularize the theory (regularization). This is of course part of the strong methodology known as (Wilsonian) Renormalization Group (RG) theory which is based on integrating out the small degrees of freedom, i.e. coarse-graining. What we "observe" as a magnetic field is the big picture, i.e., large-scale field ${\bf B}_l$ at larger scales $l$ not tiny details, i.e., field fluctuations, which can be summed over on much smaller scales. This is the gist of RG methodology which plays also a crucial role in our presentation here. As for the role of turbulence, it is well known that in both plasma physics and astrophysics, magnetized fluids of interest are usually also turbulent due to external forcing or different internal instabilities or in fact even due to reconnection itself \citep{Review2020}. Even if the system is initially non-turbulent, reconnection  can make the flow turbulent. Unlike laminar flows in ideal magnetized fluids, where magnetic field is approximately frozen into the flow (Alfv\'en flux-freezing; \cite{Alfven1942}), in turbulence, magnetic field follows the flow only in a statistical sense: this is stochastic flux-freezing formulated by Eyink \citep{Eyink2011}. Turbulent flows have non-trivial features like unpredictability, enhanced mixing and spontaneous stochasticity which tend to tangle the threading magnetic field stochastically, increasing its spatial complexity in a geometric sense \citep{JV2019}. Because of strong magnetic tension forces, at some point, the field may relax to a smoother configuration which in turn launches eruptive fluid motions \citep{Jafari2020,NanoflareAAS,Jafari2021}, potentially observable as "reconnection" events, e.g., on the solar surface.

\bibliography{main}{}

\end{document}